# The intersection of video capsule endoscopy and artificial intelligence: addressing unique challenges using machine learning

Shan Guleria[1], Benjamin Schwartz[1], Yash Sharma[2], Philip Fernandes[3], James Jablonski[2], Sodiq Adewole[2], Sanjana Srivastava[3], Fisher Rhoads[3], Michael Porter[2], Michelle Yeghyayan[3], Dylan Hyatt[3], Andrew Copland[3], Lubaina Ehsan[3], Donald Brown[4], Sana Syed[3]

1. Rush University Medical Center, Department of Internal Medicine. Chicago, IL 60607
2. University of Virginia, Systems and Information Engineering. Charlottesville, VA 22903
3. University of Virginia, Department of Pediatrics. Charlottesville, VA 22903
4. University of Virginia, Data Science Institute. Charlottesville, VA 22903

Corresponding author. Dr. Sana Syed, MD. Email: ss8xj@virginia.edu

# Abstract

Introduction: Technical burdens and time-intensive review processes limit the practical utility of video capsule endoscopy (VCE). Artificial intelligence (AI) is poised to address these limitations, but the intersection of AI and VCE reveals challenges that must first be overcome. We identified five challenges to address. Challenge #1: VCE data are stochastic and contains significant artifact. Challenge #2: VCE interpretation is cost-intensive. Challenge #3: VCE data are inherently imbalanced. Challenge #4: Existing VCE AIMLT are computationally cumbersome. Challenge #5: Clinicians are hesitant to accept AIMLT that cannot explain their process.

Methods: An anatomic landmark detection model was used to test the application of convolutional neural networks (CNNs) to the task of classifying VCE data. We also created a tool that assists in expert annotation of VCE data. We then created more elaborate models using different approaches including a multi-frame approach, a CNN based on graph representation, and a few-shot approach based on meta-learning.

Results: When used on full-length VCE footage, CNNs accurately identified anatomic landmarks (99.1%), with gradient weighted-class activation mapping showing the parts of each frame that the CNN used to make its decision. The graph CNN with weakly supervised learning (accuracy 89.9%, sensitivity of 91.1%), the few-shot model (accuracy 90.8%, precision 91.4%, sensitivity 90.9%), and the multi-frame model (accuracy 97.5%, precision 91.5%, sensitivity 94.8%) performed well.

Discussion: Each of these five challenges is addressed, in part, by one of our AI-based models. Our goal of producing high performance using lightweight models that aim to improve clinician confidence was achieved.

## Introduction

Gastrointestinal endoscopy remains the gold standard for diagnosis and, often, treatment of a wide variety of gastrointestinal diseases, including celiac disease, gastrointestinal bleeding (GIB), esophagitis, and numerous malignancies. When microscopic data is required for diagnosis, endoscopy remains unmatched. However, many GI diseases can be diagnosed with gross endoscopic findings alone, and in some cases, these disease processes may be present in areas of the GI tract that an endoscope cannot reach, such as the jejunum and proximal ileum. In such cases, video capsule endoscopy (VCE) has been shown to be effective in diagnosis while also being substantially less burdensome on the patient than standard GIE [1] indeed for small bowel bleeding in particular, VCE is an indispensable tool in diagnosis and ultimately management [2].

However, there remain significant barriers to the regular utilization of VCE in practice. As the wireless capsule must pass through the entire GI tract, a large amount of video data must be viewed by an expert endoscopist in order to make a diagnosis, resulting in a significant cost and time burden. In addition, much of the data is difficult to interpret as the capsule tumbles stochastically through the GI tract. Problems of this nature are increasingly being tackled by artificial intelligence and machine learning tools (AIMLT), and indeed such tools have already been applied to other endoscopic challenges in medicine. In particular, convolutional neural networks (CNNs) are a particular kind of AIMLT that use progressive mathematical filters to identify progressively more complex features of an image or video [3]; CNNs have been shown to be especially effective in the classification of images [4,5] and in the interpretation of video capsule endoscopy data in particular [6,7].

In fact, multiple recent investigations have utilized CNNs to tackle current challenges in VCE. Xie and colleagues trained a CNN on nearly 3000 capsule studies and found an increased detection rate of small bowel pathologic findings as well as reducing reading time by 89.3% compared to conventional human reading. However, this study was limited in that the reference standard was a heterogenous panel of VCE readers with different experiences and there was a lack of standardized, predefined reading speeds for comparison [8]. Afonso and colleagues trained a CNN with preprocessed and labeled images to identify ulcers and erosions in the small bowel; their model had a sensitivity of 90.8%, a specificity of 97.1% and an accuracy of 95.6% [9]. Another model by Ferreira and colleagues also detected ulcers with an accuracy of 92.4% [10]. While these and other studies were able to achieve admirable performance, the extensive curation and annotation of training and testing datasets and the computational power necessary for the AIMLT to run efficiently may have limited the practical applications of similar models.

The recent introduction of large language models such as ChatGPT to the general public has thrust the issue of AIMLT in medicine to the forefront of the minds of stakeholders in the field [11]. Recent studies towards AIMLT in medicine demonstrates that while attitudes towards AIMLT are variable, there are consistent concerns surrounding whether these tools actually improve clinical outcomes and workflow as well as whether such tools are able to explain to clinicians how they reach their conclusions [12,13]. Additionally, VCE data itself poses unique challenges to those who aim to interpret it. Thus, the introduction of AIMLT tools to VCE interpretation introduces a new set of challenges that must be addressed. This paper highlights five such challenges and proposes solutions in the form of novel AIMLT:

> Challenge #1: VCE data are stochastic and contains significant artifact.
> Challenge #2: VCE interpretation is cost-intensive.
> Challenge #3: VCE data are inherently imbalanced.

Challenge #4: Existing VCE AIMLT are computationally cumbersome.

Challenge #5: Clinicians are hesitant to accept AIMLT that cannot explain their process.

# Methods

Before addressing each of these challenges specifically, an overview of the methods employed in this paper is necessary. Rather than organize these methods by challenge, we will organize them according to the specific model employed for ease of reference. It is worth noting that due to variability in data acquisition, the same training and testing datasets were not used for all models. This study was approved by the University of Virginia Institutional Review Board for Health Science Research (IRB # 20650).

Annotation Tool[14]

A customized user interface was created using the Plotly Dash framework [15] with a MySQL back end to enable rapid annotation of VCE recordings. Annotators could select abnormalities from a radio-button list and view multiple frames simultaneously, facilitating contextual cues and rapid labeling. Code for creating this app is available at the following link: https://github.com/SyedLab-GI/VCEapp.

Anatomic Landmark Detection Models [16]

We compared the performance of four CNNs in their ability to detect anatomic landmarks within the GI tract: Visual Geometry Group (VGG)Net [17], GoogLeNet [18], Residual Network (ResNet) [19], and AlexNet [4]. Our dataset consisted of approximately 200,000 frames of VCE videos from nine patients, each video being assessed in its entirety and including the four main anatomical regions: esophagus, stomach, small bowel, and colon. To address the significant class imbalance in the dataset, with over 80% of frames capturing images of only the small bowel, we employed up-sampling techniques. Images from other regions (esophagus, stomach, and colon) were up-sampled to balance the class distribution as the majority of images were from the longest part of the GI tract, the small bowel. Additionally, we applied random rotations and cropping to generate additional training examples for these regions. The augmentation process was performed exclusively on the training set to balance the classes, while the test set remained unaltered. For each model, we utilized gradient weighted-class activation mapping (Grad-CAM) [20] to visualize the learned features and regions of importance, the code for which is publicly available here: https://github.com/SyedLab-GI/GradMix.

Multi-Frame Model [14]

The dataset used for training consisted of 190,768 frames from 15 patient recordings with the SB3 PillCam. Four annotators labeled the data using the customized annotation system, and two physicians provided quality control. The data were divided into five classes: Small Bowel Normal, Stomach, Colon, and then a fifth class called "small bowel abnormal" which consisted of ulcers (557 frames), clot (403 frames), masses (1874 frames), polyps (270 frames), angioectasia (581 frames), edematous villi (8004 frames), bleeding (1836 frames), and other (1415 frames). Importantly, this labeling was done at the frame level. The evaluation included both localization and abnormality detection tasks on a single model. Performance metrics for this model pertain only to this small bowel "abnormal" class due to the importance of this task in the clinical setting as compared to evaluations of the stomach or colon. Video prediction time estimates were generated by predicting 20,000 frames to simulate an entire video being processed.

A late fusion approach was chosen, where frames are merged at the top of the network, to incorporate adjacent frames into predictions. Multiple models with varying levels of adjacent frames were constructed, with a base consisting of the convolutional layers and global max-pooling layer of a pre-trained MobileNet [21]. The features from these base columns were concatenated, and a fully-connected layer with a Softmax prediction layer was added.

Graph CNN Model [22]

VCE data from five patients were used to fine-tune the pre-trained VGG-19 network [17], and subsequently nine VCE videos were used for the testing component. The dataset was annotated in similar fashion to the other models.

An automatic temporal segmentation technique based on change point detection was employed to divide videos into uniform and identifiable segments[23–28]. A Graph Convolutional Neural Network model, specifically based on the GraphSage architecture [29], was then utilized to learn representations of each video segment. By leveraging topological relationships between frames in the video, this CNN model captures the spatial, temporal, and neighborhood information. Importantly, this architecture necessitates that the model be tested on segments of VCE data from the same video in order for it to accurately assess those relationships. The model's goal was to classify frames as either normal or abnormal, with the abnormalities including angioectasias, bleeding, erosions, erythema, and ulcers.

The Graph CNN model incorporates an attention mechanism to identify the most informative frame in the sequence. After applying an aggregator function to aggregate the weighted features, the final layer of the network was replaced with a temporal pooling layer, which facilitates the localization of relevant abnormal frames within each abnormal video segment.

The pretrained VGG-19 network, initially trained on the large ImageNet dataset, was fine-tuned on five VCE videos to capture and learn representations of abnormalities present in the frames. Weighted oversampling was applied during the model's fine-tuning process to balance exposure to different classes.

Meta-Learning and Few-Shot Learning Models [30]

Standard supervised Convolutional Neural Networks (CNN) classifiers require a large number of labeled examples for training and are sample inefficient. In contrast, humans can learn new concepts from just a few instances and quickly generalize[31]. To mimic this ability, meta-learning was employed to train a CNN classifier on certain lesion categories and then evaluate its performance on a different set of lesion categories using few-shot learning.

Unlike other models, the goal with this model was to train a model off a relatively small dataset that emphasizes pathology. Therefore, this dataset included data from 52 VCE videos with select frames chosen to represent four pathology categories: Whipples, ulcer, bleeding, and angioectasia. The result was 5,360 frames with an average of 1,072 for each of the four categories. The data was split into train-test sets using 70% for the training set and 30% for the test set. A feature extractor was trained using meta-learning to learn representations for different small bowel lesions. While four categories were used for the training set, the testing set included a fifth category on which the model had not been trained, and the model had to infer the existence of this fifth category as being distinct from the other four. Anywhere from one to nine supporting examples were provided to the model to assist in the categorization of this fifth group of lesions, and these supporting examples represent the "few shots" for which the model is named.

# Results

Challenge #1: VCE data are stochastic and contains significant artifact.

Poor image quality and artifact are significant barriers to reading VCE in clinical practice [32]. Such artifact-laden datasets are particularly challenging for traditional machine learning tools that have trouble generalizing features across limited datasets [33], but we hypothesized that a high level of accuracy could be achieved with CNNs. Indeed, our data show that multiple different CNNs can accurately identify the esophagus, stomach, small bowel, and colon as distinct entities (Table 1), and all four of the models analyzed were able to achieve similar performance metrics when identifying these anatomical landmarks from VCE data (Table 2), with the highest accuracy at 99.1%. Importantly, these CNNs were able to achieve such results using a dataset consisting of full-length VCE footage from real patients, as opposed to datasets showing only idealized healthy and pathologic tissue.

Challenge #2: VCE interpretation is cost-intensive.

In part due to the significant artifact as noted above, VCE data is cumbersome and cost-intensive for gastroenterologists to review and interpret. Although AIMLTs have the potential to improve this issue, standard supervised CNNs require large quantities of training data that are also annotated so the CNN can learn to categorize new datapoints [31]. In the context of VCE AIMLT, this would require that an expert in VCE spend significant time labeling and annotating frames of VCE data in order to train models, which is even more time- and cost-intensive than interpreting VCE in clinical practice.

Most open-source video annotation tools are focused on object tracking in frame-by-frame video, but in VCE the primary concern is identifying whether or not disease exists in a given frame, both for the purposes of CNN training and for clinical utility [34]. As such, we developed our own VCE annotation mechanism to partially address the issue of time-intensive VCE review (Figure 1) [14]. Although direct comparisons to other annotation mechanisms were not performed, lab annotators estimated that time spent annotating VCE recordings decreased from approximately 150 minutes to 45 minutes, around a 70% reduction.

However, despite the successes of this annotation tool, annotating and labeling VCE data remains costly, and the majority of AIMLT rely heavily on a large, annotated training dataset. To combat this challenge, we developed a novel AIMLT using a graph CNN model, which was trained in a "weakly supervised" fashion. In other words, the model was able to learn about frame-level features while being trained on data that was labeled at the level of short video segments rather than frame-by-frame. The results are summarized in Table 3 which show that when tested on some VCE footage, accuracy was as high as 89.9% with a sensitivity of 91.1%. However, when applied to other VCE videos, performance was substantially lower (e.g. 56% accuracy for one video in Table 3).

Additionally, we aimed to further reduce the annotation burden and developed an AIMLT based on state-of-the-art machine learning techniques called meta-learning and few-shot learning, which allow the model to learn to identify features of VCE data, classify the frames into categories, and generalize this knowledge to new tasks [30]. The result is a prediction model that can be trained on a relatively small number of annotated images that show disease processes but is still capable of classifying new, never-before-seen lesions. Table 4 summarizes the results of our few-shot learning model. Across the five different categories of lesions presented to the three CNNs, AlexNet in particular achieved the highest performance with an accuracy of 90.8%, a precision of 91.4%, and a sensitivity of 90.9% (Table 4).

Challenge #3: VCE data are inherently imbalanced.

The performance of AIMLT relies heavily on the data on which they are trained, and in particular these tools tend to have difficulty with imbalanced datasets [33]. Unfortunately, biomedical data are often inherently imbalanced due to the distribution of normal and abnormal datapoints [35,36] and VCE data are no exception. For example, in terms of the distribution of anatomical landmarks, the vast majority of VCE frames in a given patient are obtained from the small bowel as it is by far the longest segment of the GI tract. Additionally, a given patient may have tens of thousands of frames in their VCE footage, but the small bleeding ulcer in their jejunum may only make up one or two of those frames. Thus, a typical series of VCE images from a given patient or group of patients results in imbalanced data on which it is difficult to train AIMLT. Other studies have utilized well-curated, annotated VCE datasets to solve this issue [37–39], but these datasets are not representative of real patients and are costly to produce and maintain.

We propose multiple solutions to this issue. First, we employed data up-sampling as outlined above in the methods in order to better balance the datasets and prevent small bowel images from being over-represented. Additionally, the novel meta-learning and few-shot learning AIMLT described above are not only capable of accurately classifying the pathologies on which they were trained, but they are also capable of categorizing new categories of pathology that they had never seen before. The result is an AIMLT that does not require large amounts of annotated data and is not adversely affected by the fact that VCE data consists overwhelmingly of normal tissue.

Challenge #4: Existing VCE AIMLT are computationally cumbersome.

Given the amount of time it takes for a capsule to move through the GI tract, VCE footage can generate an extensive dataset of over 50,000 image frames per patient. The sheer volume of images generated by VCE poses a computational challenge when AIMLTs are used for their analysis. Analyzing each frame of the video and extracting relevant features for disease identification requires significant computational resources and time.

The clinical application of these methods can be limited by the availability of the graphics processing units (GPUs) necessary for processing such a large amount of data. GPUs are utilized to increase the computing speed and efficiency of deep learning algorithms, but the more complex the network, the more robust GPU necessary for the model to complete its task in a reasonable amount of time. For example, the current state-of-the-art CNN-based models such as the one presented by Ding *et al.* demonstrates impressive performance in identifying abnormalities in small-bowel capsule endoscopies [40]. However, the model's architecture and complexity demand significant computational power and resources making it much less practical for real-time clinical application. ResNet152 has 58.3 million parameters in its architecture; in contrast, our proposed multi-frame models have parameters ranging from 4.7 to 7.3 million and require relatively less GPU memory [14]. Despite being more lightweight, when comparing the ResNet152-based model's performance to our multi-frame models, our m1110 model is superior in terms of accuracy (97.5% vs 96.6%), precision (91.5% vs 90.3%), and sensitivity (94.8% vs 91.1%) on the task of identifying abnormalities in the small bowel (Table 5).

Challenge #5: Clinicians are hesitant to accept AIMLT that cannot explain their process.

Since the advent of AIMLTs in medicine, there has been concern regarding the so-called "black box" of artificial intelligence, which describes the lack of human understanding of how exactly the AIMLT arrives at its prediction or conclusion [41]. Particularly when such tools are applied to diagnoses such as gastrointestinal malignancy or acute bleeding requiring intervention, this black box poses challenges to practical utilization.

A proposed solution to this challenge is the Gradient-weighted Class Activation Map (Grad-CAM), which visually highlight regions of images that are relevant to the classification of that image [20], thus shedding some light on the black box of AIMLTs. Grad-CAMs have been employed to explain AIMLT decision-making in other medical media, such as pathology slides and confocal imaging [42], but to our knowledge Grad-CAMs applied to VCE data are less common, and those that do exist are built on meticulously balanced and annotated datasets [43]. Not only did our CNN-based AIMLTs successfully distinguish anatomic landmarks within the GI tract as described above, Grad-CAMs from these models show that they successfully ignore artifact such as bubbles in the esophagus and focus on features of the tissue itself (e.g. villi in the small bowel).

Additionally, we developed a novel AIMLT to analyze VCE that takes inspiration from the methods employed by trained human endoscopists. As it stands, current state-of-the-art AIMLTs make predictions based on single frames of VCE data, and as explained, the majority of published results focus on cleanly labeled examples isolated from full-length videos. In stark contrast, human endoscopists reading VCE studies rely on contextual cues to identify abnormalities in VCE data, scanning between adjacent frames to identify patterns [44], particularly for larger lesions such as serpiginous ulcers. We therefore designed two novel AIMLTs that drew inspiration from human endoscopists in their approaches to image recognition. The graph CNN model described above captures relationship information from features in each frame in order to relate each frame to its neighbors. While this approach is novel and valuable, it is not inherently intuitive to an observer what exactly the model is doing to reach this understanding of relationships between frames. We therefore developed a model that relies on a multi-frame approach, incorporating the features of adjacent frames into its predictions for any given frame, and we set it to the task of identifying abnormalities in images of the small bowel. These results (Table 5) demonstrate that incorporating adjacent frames in VCE modeling improved performance in both localization and abnormality detection tasks compared to current state-of-the-art models (represented by the ResNet152 model), all without substantially increasing prediction time or computational requirements. In particular, the multi-frame approach led to a 3.7% increase in sensitivity, a key metric for AIMLT designed not to miss important abnormalities.

## Discussion/Conclusion

Our work focused on addressing these complex challenges at every level, both for the purposes of advancing research in AIMLT as they pertain to VCE as well as for practical clinical use cases. Clinicians in their practice may spend over an hour reading VCE images, and experts delicately annotating images for CNN training are likely to spend even more time. The annotation tool stands to assist with the labeling of VCE datasets, allowing for improvements in supervised CNNs developed in our lab and also other labs, as we have made the tool public for all to use. Future work involving our VCE annotation tool will demonstrate exactly how much time can be saved using this tool both for CNN training and for clinicians hoping to improve the efficiency and accuracy of their own VCE review. For practical purposes, any AIMLT that are able to achieve excellent sensitivity (i.e. very few false negatives, known in computer science as "recall") at the expense of specificity (i.e. higher false positive rate) may still prove extremely useful to the practicing gastroenterologist. With a low proportion or prevalence of pathology per frame, an increasingly sensitive test will provide a high negative predictive value as well, making it a powerful clinical tool. For example, such an AIMLT may be able to highlight a small fraction of VCE data that is most likely to contain pathology, as clinically important pathology may only be visualized on a few frames out of tens of thousands of frames in a single study; this would significantly reduce the burden of reviewing VCE footage for the gastroenterologist, as VCE studies can take more than 45-60 minutes to adequately review [44]. Reducing reading times may theoretically also increase overall sensitivity of pathology by reducing clinician fatigue and cognitive workload. In such real-world scenarios, AIMLT that are computationally expensive would be quite limited in their clinical applications; therefore, we created models that are accurate and also computationally lightweight, demonstrating that AIMLT applied to VCE can perform well while being practical.

Another goal of our work was to address the so-called "black box" of AIMLT, which describes the inability of lay observers and often the model creators themselves to explain how exactly AIMLT arrive at their conclusions or predictions. In light of the recent adoption of large language models in both daily life in general and the medical field in particular, it is clear anecdotally and in the studies described above that physicians and patients alike are uncomfortable with the introduction of AIMLT that lack "explainability." We aimed to address this in our models by incorporating visual representations of areas of interest for the model in the form of Grad-CAMs. Perhaps more interestingly, we also drew inspiration from human analysis of VCE data. Our multi-frame approach incorporates data adjacent to each frame in the feature detection process, which imitates the way that a human gastroenterologist scrolls back and forth to understand what is happening in each frame. Our results demonstrate that such an approach improves accuracy and, importantly, sensitivity compared to single-frame approaches without significantly increasing the time the AIMLT takes to achieve these results. At an even higher level, our graph CNN model is able to map out the relationships, similarities, and differences between features of each image, mimicking the natural intuition of a human gastroenterologist, who would instinctively understand the relationship between, for example, the red color and twisted shape of a serpiginous bleeding ulcer. However, because the graph CNN is highly dependent on the quality of the VCE data, its results vary widely depending on the specific data on which it is tested. Further research with larger datasets will be necessary to determine which aspects of the VCE data result in lower performance.

Our group also set out to create novel AIMLTs for VCE interpretation specifically on full-length VCE video data in order to more closely approximate real-world conditions. As explained, this created new challenges as real-world VCE data is highly imbalanced, filled with artifact, and contains relatively small amounts of pathology. Our novel few-shot learning and meta-learning approaches to these challenges was inspired by work being done on aerial imaging [31]. Aerial images taken by drones are quite similar to VCE images; both contain vast swaths of relatively homogeneous images and are cumbersome for humans to individually review and annotate, but both also can contain life-saving information hidden within a sea of normal images. Applying these novel techniques to VCE data resulted in a model that can not only categorize VCE images accurately with relatively little human annotation in the training process but also one that can learn to identify new categories on which it was not specifically trained. Such a model can be invaluable in endoscopic images as pathologies such as small bowel tumors are exceedingly rare and may not be found in many training datasets but are nonetheless important for AIMLT in VCE to be able to identify. The few-shot model also addresses an important drawback of both the multi-frame and graph CNNs, which required extensive labeling for training purposes.

Areas for continued AIMLT improvement and further research remain, however. For example, our data consists of a relatively small number of recordings from a single capsule manufacturer, and our annotations for the training data were based on the consensus of relatively few annotators. More extensive experiments with larger datasets using a wide variety of capsule models and annotators will be required to better evaluate the accuracy of these AIMLT. Additionally, future work in this field may aim to provide not only predictions of disease states but also high accuracy regarding such predictions, as clinician trust of AIMLT is a major issue. Finally, if the goal is a practical use of AIMLT that can improve clinical practice, future studies should explore whether clinicians can achieve similarly accurate results and save time, which our study did not examine specifically.

In conclusion, VCE data present significant challenges both to practicing physicians as well as to data scientists attempting to use AIMLT to interpret VCE data. We employed multiple novel data science techniques to address these challenges. In doing so, we developed (1) a tool with which additional high-quality data can be obtained for further model building, (2) accurate and lightweight AIMLT for predicting both anatomic landmarks and disease states despite the challenges of VCE data, and (3) AIMLT specifically aimed at increasing clinician confidence in machine learning. Our work highlights the importance of ensuring that AIMLTs are not only accurate but also practical for clinical use in the near future.

# Figures and Tables

Tables

**Table 1.** Normalized confusion matrix of the CNNs used to identify anatomical landmarks in VCE images. Abbreviations: Visual Geometry Group (VGG), Residual Network (ResNet).

| True-Label | Model | Predicted Label | | | |
|---|---|---|---|---|---|
| | | Esophagus | Stomach | Small Bowel | Colon |
| Esophagus | VGGNet | 0.990 | 0.000 | 0.009 | 0.001 |
| | ResNet | 0.995 | 0.000 | 0.003 | 0.002 |
| | GoogLeNet | 0.957 | 0.001 | 0.012 | 0.031 |
| | AlexNet | 0.979 | 0.000 | 0.019 | 0.002 |
| Stomach | VGGNet | 0.000 | 0.926 | 0.000 | 0.074 |
| | ResNet | 0.000 | 0.907 | 0.000 | 0.093 |
| | GoogLeNet | 0.000 | 0.963 | 0.000 | 0.037 |
| | AlexNet | 0.000 | 0.870 | 0.000 | 0.130 |
| Small bowel | VGGNet | 0.004 | 0.000 | 0.991 | 0.005 |
| | ResNet | 0.070 | 0.003 | 0.848 | 0.079 |
| | GoogLeNet | 0.072 | 0.006 | 0.825 | 0.097 |
| | AlexNet | 0.019 | 0.000 | 0.972 | 0.009 |
| Colon | VGGNet | 0.001 | 0.002 | 0.013 | 0.984 |
| | ResNet | 0.004 | 0.004 | 0.001 | 0.991 |
| | GoogLeNet | 0.008 | 0.008 | 0.003 | 0.981 |
| | AlexNet | 0.003 | 0.004 | 0.017 | 0.977 |

**Table 2.** Comparative performance of the CNNs used to identify anatomical landmarks in VCE images. Performance metrics are defined as below. Abbreviations: Visual Geometry Group (VGG), Residual Network (ResNet).

| Model | Accuracy | Precision | Sensitivity | F-score |
|---|---|---|---|---|
| **VGGNet** | 99.1% | 93.5% | 97.3% | 95.3% |
| **ResNet** | 87.8% | 61.9% | 93.5% | 70.4% |
| **GoogleNet** | 85.4% | 57.0% | 93.1% | 64.2% |
| **AlexNet** | 97.3% | 88.7% | 95.0% | 91.6% |

$$Sensivity = Recall = \frac{TP}{(TP + FN)}$$

$$Accuracy = \frac{(TP + TN)}{(TP + FP + FN + TN)}$$

$$Precision = PPV = \frac{TP}{(TP + FP)}$$

$$F = 2 * \frac{Precision * Recall}{(Precision + Recall)}$$

TP = True Positive
FP = False Positive
TN = True Negative
FN = False Negative
PPV = Positive Predictive Value
F = F-score

**Table 3.** Comparative performance of the graph CNN model in identifying normal versus abnormal frames in the four VCE videos in the testing set.

| Metrics | Video 6 | Video 7 | Video 8 | Video 9 |
|---|---|---|---|---|
| Frames | 14,173 | 16,909 | 10,037 | 19,104 |
| Segments | 770 | 1124 | 248 | 1071 |
| Disease Categories | 5 | 2 | 3 | 3 |
| **Precision** | 82.0% | 69.8% | 33.6% | 76.7% |
| **Sensitivity** | 91.1% | 80.4% | 60.1% | 88.9% |
| **F-score** | 89.9% | 82.2% | 57.8% | 85.9% |
| **Accuracy** | 89.9% | 84.8% | 56.0% | 85.9% |

**Table 4.** Comparative performance of three different CNNs, each with the addition of few-shot learning layers, in their ability to identify four known categories of pathology in VCE footage (Whipples, ulcer, bleeding, and angioectasia) as well as a fifth unknown category. Abbreviations: Visual Geometry Group (VGG), Residual Network (ResNet).

| Model | Accuracy | Precision | Sensitivity | F-score |
|---|---|---|---|---|
| **VGGNet + Few Shot** | 89.3% | 89.5% | 88.8% | 89.0% |
| **ResNet + Few Shot** | 85.4% | 82.3% | 82.4% | 82.4% |
| **AlexNet + Few Shot** | 90.8% | 91.4% | 90.9% | 91.0% |

**Table 5.** Comparative performance of three different CNNs in the identification of normal versus abnormal frames in VCE footage. The Residual Network (ResNet152) model incorporates only the frame in question into its prediction. The m0110 multi-frame model relies on the frame in question plus the frame immediately preceding it. The m1110 multi-frame model incorporates the frame in question and the two preceding frames.

| Model | Accuracy | Precision | Sensitivity | F1 score | Video Prediction Time (minutes) |
|---|---|---|---|---|---|
| **ResNet152** | 96.6% | 90.3% | 91.1% | 90.7% | 2.0 |
| **m0110** | 97.1% | 91.0% | 93.5% | 92.3% | 1.7 |
| **m1110** | 97.5% | 91.5% | 94.8% | 93.1% | 2.2 |

Figures

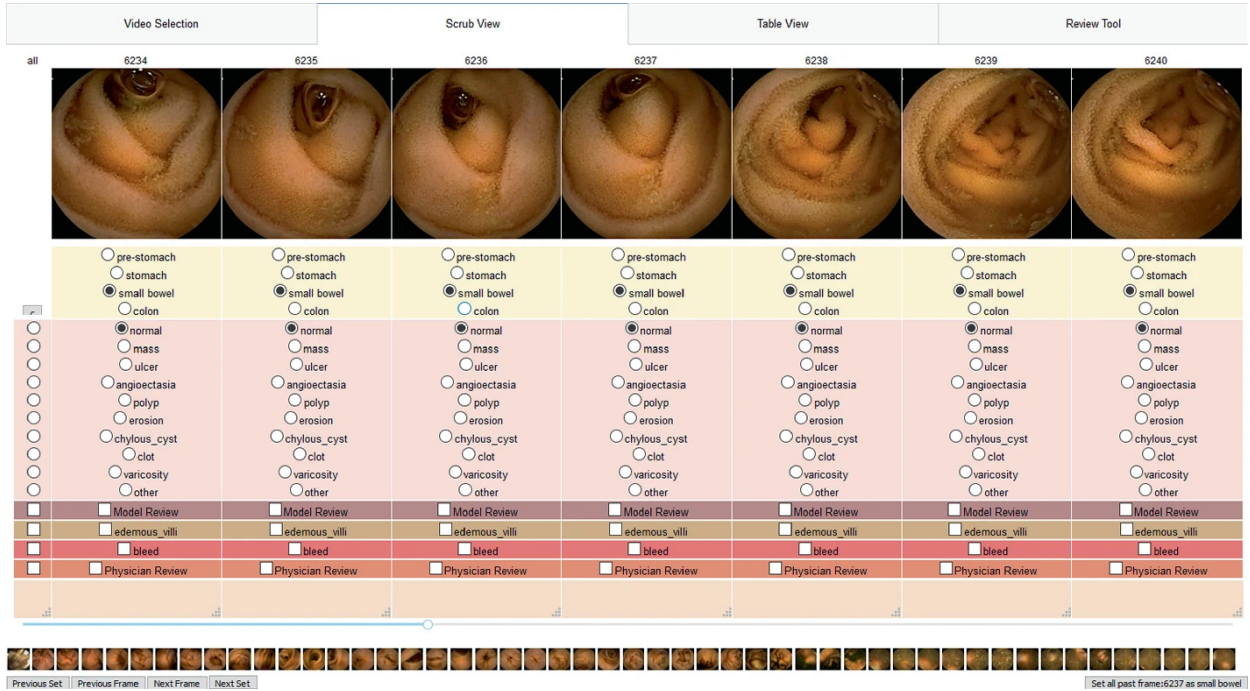

**Figure 1.** Annotation tool interface to standardize frame labeling. A multi-frame display was chosen to enable contextual understanding and improve labeling speed. Shown here are small bowel images, all of which are normal. Abnormalities are subdivided into categories with special options for additional review or bleeding for easier subsequent identification.

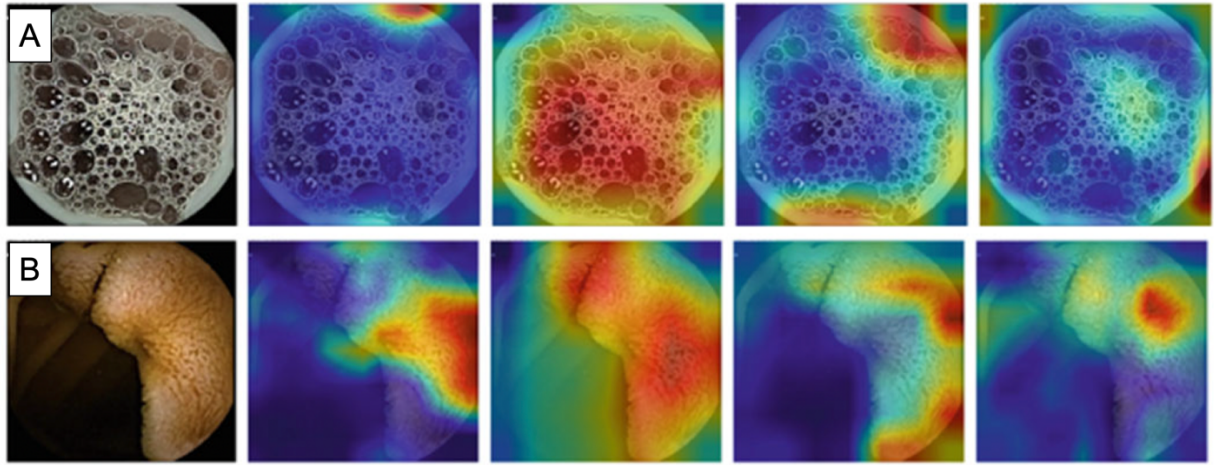

**Figure 2A.** Gradient weighted-class activation mapping (Grad-CAM) analysis of the esophagus. From left to right: no filter, AlexNet, GoogLeNet, Residual Network (ResNet), Visual Geometry Group (VGG). GoogLeNet notably focuses on diffuse bubbles as an identifying feature of the esophagus.

**Figure 2B.** GradCAM analysis of the small bowel. From left to right: no filter, AlexNet, GoogLeNet, ResNet-50, VGG. All four networks visually recognized the villi of the small bowel mucosa as identifying features to distinguish it from other landmarks.